# Spatial Light Modulator for wavefront correction

A. Vyas[1,2], M. B. Roopashree[1], Ravinder Kumar Banyal[1], B Raghavendra Prasad[1]

[1]Indian Institute of Astrophysics, Bangalore-560034, India

[2]Indian Institute of Science, Bangalore-560012, India

**Abstract:**

*We present a liquid crystal method of correcting the phase of an aberrated wavefront using a spatial light modulator. A simple and efficient lab model has been demonstrated for wavefront correction. The crux of a wavefront correcting system in an adaptive optics system lies in the speed and the image quality that can be achieved. The speeds and the accuracy of wavefront representation using Zernike polynomials have been presented using a very fast method of computation.*

## 1. INTRODUCTION

The possibility of compensating wavefront aberrations was first suggested by Babcock [1]. The idea of compensation given by him was based on the electrostatic forces acting on oil film. The design was that the refractive index changes on the oil film can compensate for the aberrations. Later on when the proposal of Hartmann was implemented by Shack and Platt [2], to make the first Shack Hartmann sensor, people thought of compensating the wavefront using optics (deformable mirrors) rather than refractive index changes of materials. After the development of liquid crystals, the possibility of using the changing refractive index property for phase restoration applications has been explored. Most applications where phase restorations become essential are human eye aberration corrections and atmospheric turbulence compensations. The use of Spatial Light Modulator (SLM) as a wavefront corrector has been demonstrated earlier [3]. The advantage of using Liquid Crystal Spatial Light Modulator (LCSLM) is that it can be used as a wavefront sensor and corrector as well [4]. Attempts are being made to retrieve phase from the diffraction pattern of an aperture array [5]. Phase only corrections using SLM have been done in the earlier studies [6]. The attention on SLM is because of the various advantages it can offer which include high resolution, low cost and compactness. If it is possible to overcome some drawbacks like the need of monochromatic polarized light and relatively low response time, it is possible to achieve good results with this device. A simple method for phase modulating unpolarized light with a double pass through a nematic liquid crystal was described earlier [7]. The broadband performance of a polarization-insensitive liquid crystal phase modulator has been proposed and its effects in adaptive system have been quantified [8]. More research on how this device can be used with unpolarized broadband light can make LCSLMs the best choice for adaptive optics.

Use of Zernike polynomials for describing aberrations of an optical system is well accepted. Since Zernike moments outperform any other moments in representation of images, they are the best for aberration quantification, if known how to use them correctly. The need for quantification of wavefront aberration requires us to compute the Zernike moments which weigh each of the Zernike polynomials. Many algorithms exist for computation of these moments. The mathematical complexity of the definition of the Zernike polynomials will not allow fast computation. Recurrence relationships have been developed by some authors like Prata, Kintner, and Belkasim for reducing the computational time [10, 11]. Coefficient method and the q-recursive method also increased processing speeds to a great extent [12, 13]. Each of these methods has their advantages and disadvantages. A hybrid method involving all the above methods has been developed to increase the speed of computations [14]. Very recently, Hosny proposed a fast and accurate method of computing Zernike moments [15]. This method minimizes the geometrical errors by a proper mapping of the image and the approximation errors are removed by using exact Zernike moments for reconstruction of the image using Zernike polynomials.

This paper details the generation and correction of Zernike aberrations imposed on an image using two SLMs. We present the production of different Zernike aberrations using the LC2002 Holoeye SLM. Wavefront correction is achieved by imposing phase corrections on the second SLM which is a Meadowlark Optics Hex127 SLM. Both the SLMs were characterized in terms of their nonlinearity and phase retardance for best performance. The phase compatibility of the SLMs is checked and is presented here. To obtain the real time closed loop wavefront correction, one needs to quantify the aberration. For the quantification of aberrations, we need to read the image and the aberrations involved in terms of the Zernike moments. We present the results of the calculation of Zernike moments. If the aberrations were to be represented by the derivatives of Zernike polynomials, some prior knowledge of the aberrations becomes very important for the correction of the aberrations. If one tries to correct a lower order aberration using higher orders, absurd results will be obtained. It is very important to have an idea of the number of orders of Zernike polynomials to be used for representing aberrations. To obtain exact Zernike moments, iterations have to be performed, which will make it difficult to estimate the timescales of computations. If the speeds are not an issue, it is possible to use Zernike polynomials instead of their derivatives for the reconstruction of an aberration. We have suggested the number of orders of Zernike polynomials that must be used for aberration representation. Experiments were performed and the images were captured before and after the correction for comparison. A complete study of the effect of various aberrations and the images as they look after the imposition of Zernike aberrations has been presented. The extent of correction has been quantified and the results have been presented.

## 2. THEORY

### 2.1 Calculation of Zernike Moments

Zernike polynomials are continuous orthogonal circular polynomials defined over the unit disk. Since they form a complete set of orthogonal polynomials, any 2D function can be represented as a proper linear combination of this basis set. The Zernike polynomials are defined as,

Even Zernike polynomials:

$$Z_n^m(\rho, \varphi) = R_n^m(\rho) \cos(m\varphi)$$

Odd Zernike polynomials:

$$Z_n^m(\rho, \varphi) = R_n^{-m}(\rho) \sin(m\varphi) \tag{1}$$

where,

$$R_n^m = \sum_{k=0}^{(n-m)/2} \frac{(-1)^k (n-k)!}{k! \left(\frac{(n+m)}{2} - k\right)! \left(\frac{n-m}{2} - k\right)!} \rho^{n-2k} \tag{2}$$

$$f(x, y) = \sum_{n=0}^{M} \sum_{\substack{m \\ n-m=even \\ -n \leq m \leq n}} V_n^m R_n^m(x, y) \tag{3}$$

$$V_n^m = \begin{cases} \dfrac{n+1}{\pi} \displaystyle\sum_{\substack{k=|m| \\ n-k=even}} I_{n|m|k} \mathcal{R}_{km}, & n \neq m \\ \dfrac{n+1}{\pi} \mathcal{R}_{nm}, & n = m \end{cases} \tag{4}$$

where,

$$\mathcal{R}_{km} = \sum_{j=0}^{N} \sum_{q=0}^{\ } |m| h^q D(s, j) D(m, q) G_{k-2j-q, 2j+q} \tag{5}$$

$$s = \frac{k - |m|}{2}, \quad h = \begin{cases} -\hat{\imath}, & m > 0 \\ \hat{\imath}, & m \leq 0 \end{cases}, \quad \hat{\imath} = \sqrt{-1} \tag{6}$$

also,

$$\left.\begin{array}{l}B_{nnn} = 1 \\ B_{n(m-2)n} = \dfrac{n+m}{n-m+2} B_{nmn} \\ B_{nm(k-2)} = -\dfrac{(k+m)(k-m)}{(n+k)(n-k+2)} B_{nmk}\end{array}\right\} \quad (7)$$

The recurrence relations used for computation of factorial terms are modified for computational ease,

$$\left.\begin{array}{l}D(n,k) = \dfrac{n}{n-k} D(n-1,k) \ ; \ n \neq k \\ D(n,k) = \dfrac{1}{k} D(n,k-1) \ \ \ ; \ n = k\end{array}\right\} \quad (8)$$

The parameters V, B and D defined in equations 4, 7 and 8 are calculated beforehand since they do not depend on the aberration function.

The geometric moments are calculated using

$$G_n^m = \sum_{i=1}^{N} \sum_{j=1}^{N} I_n(i) I_m(j) f(x_i, y_j) \quad (9)$$

where,

$$\left.\begin{array}{l}I_n(i) = \dfrac{1}{n+1} \left[U_{i+1}^{n+1} - U_i^{n+1}\right] \\ I_m(j) = \dfrac{1}{m+1} \left[U_{j+1}^{m+1} - U_{j+1}^{m+1}\right]\end{array}\right\} \quad (10)$$

where,

$$U_i = \frac{2(i-1) - N}{N\sqrt{2}} \quad (11)$$

Since the parameter I does not depend on the function $f(x, y)$, it is stored in a vector form to improve speed. 'N' represents the number of moments used to Zernike moments used to represent an image.

## 2.2 Calculation of Zernike Moments

For the evaluation of the image quality, we calculated the Peak Signal to Noise Ratio (PSNR) which is related to the Root Mean Squared (RMS) error by the relation,

$$\text{PSNR} = 10 \times \log\left(\frac{2^b - 1}{\text{RMS}^2}\right) \quad (12)$$

for an b-bit image. Comparison was done between equi-sized images.

The ratio of the intensity at the Gaussian image point in the presence of aberration to the intensity that would be obtained if no aberration were present is called the Strehl ratio. Strehl ratio (S.R) is calculated using the Zernike moments of the image.

$$S.R = \sum_{n=1}^{m} \sum_{m=-n}^{n} \frac{(V_n^m)^2}{2(n+1)} \quad (13)$$

# 3. EXPERIMENTAL SETUP

The layout of the experiment is shown in Fig. 1. SLM1 is an electrically addressed LC2002 model from HoloEye Photonics. It consists of a twisted nematic LC panel (Sony LC- X016AL-6) with an active area of 26.6×20.0 mm$^2$ having 832×624 pixels. Image parameters like brightness and contrast settings can be controlled using the driver software over a range from 0-255 (grayscales). SLM2 is a Meadowlark Optics Hex127 SLM with 127 hexagonal pixels. The size of each of the pixels is 1mm. Pixels are arranged such that the effective area of the SLM forms a circular shape. We used a 15mW He-Ne laser operating at 632.8nm. The output of the laser is spatially filtered using a Newport 3-axis spatial filter mount setup. The filter consists of a 40X beam expander and a 5μm pinhole. GP is a Glan polarizer and P1, P2 are linear sheet polarizers. L1 is a doublet of focal length 20cm and L2, L3 are triplets of focal length 12.5cm and the reimaging lens L4 is a doublet of focal length 25 cm. SLM2 is at the image plane of the SLM1.

The object was placed before SLM1 and the CCD was placed on the image plane of the object. SLM1 was used for the projection of the aberration, the Zernike polynomials in our case. SLM2 was used for phase correction.

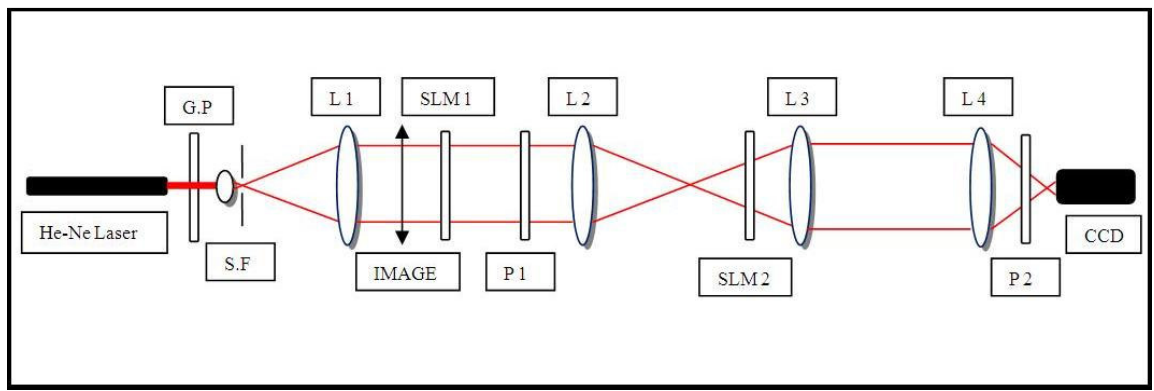

**Fig. 1. The experimental setup for wavefront correction**

# 3. RESULTS

## 4.1 Computational experiments

Zernike Moments were calculated for different 2-D discretely continuous and bounded functions and the images were reconstructed using the Zernike polynomials. The original and the reconstructed images are shown in Fig. 2. Reconstruction of Lena image has been shown in Fig.3. The PSNR has been calculated for reconstructed images using different number of Zernike moments and has been plotted in Fig. 4 and Fig. 5.

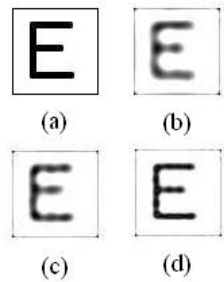

**Fig. 2. `E' image used for reconstruction. Image quality improves by increasing the number of**

**Zernike moments (a) original (b) 20 orders (c) 30 orders (d) 40 orders.**

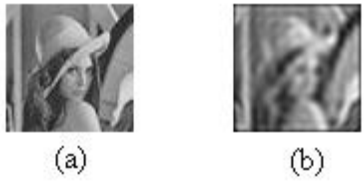

**Fig. 3. 64×64 Lena's Image (a) Original (b) Mathematically reconstructed**

**using 44 orders of Zernike polynomials.**

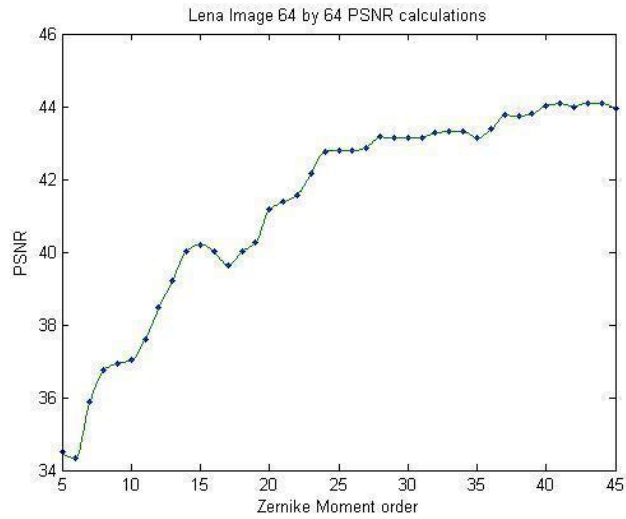

**Fig. 4. The PSNR calculations of a 64×64 image of Lena.**

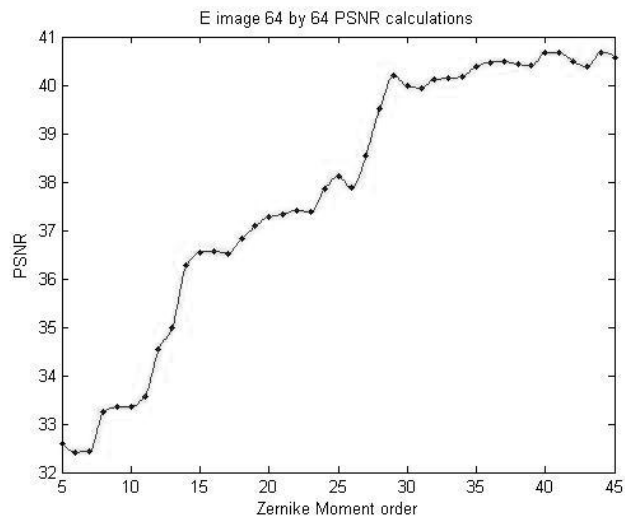

**Fig. 5. PSNR calculations of a 64×64 'E' image**

The speed at which we computed Zernike moments for 7 orders is nearly 7μs. Speeds achieved for a 64×64 image is precisely given in Table 1 and are also plotted in Fig. 6.

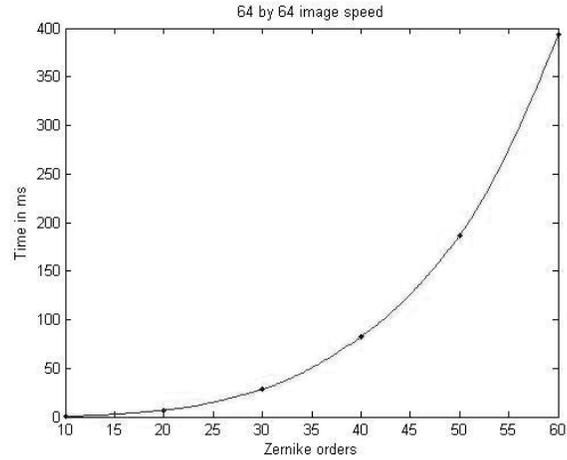

**Fig. 6. Processing time for a grayscale 64×64 image**

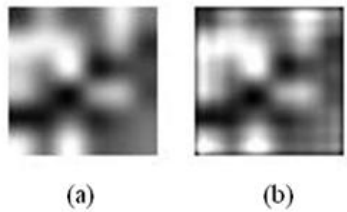

**Figure 7 Random phase image reconstruction using Zernikes (a) Original phase image**

**generated using MATLAB. Reconstruction using (b) 15 orders**

In understanding the phase disturbed images of low resolution, especially for astronomical applications; we took random images of 36 degrees of freedom, i.e. 36 phase values and reconstructed those using Zernike moments as shown in Fig. 7. In Fig. 8, we plotted the image quality (normalized PSNR) as a function of the number of orders of Zernike polynomials used for the reconstruction. The optimum value of Zernike orders to be used for image reconstruction depends on two factors, the accuracy of representation that the application demands and the speed at which the processing has to be done. This graph suggests that for low resolution images like this one, using less number of orders can give very good results unlike for Lena's image, where using 45 orders is also not satisfactory.

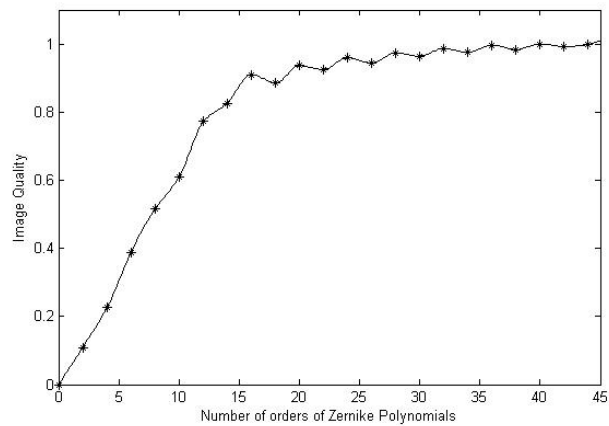

**Fig. 8. Image quality with the number of orders for a random phase image**

**Table 1. Processing speed improvement for a 64×64 image**

| Zernike Orders | Earlier Reported [15] $t_1(ms)$ | Speeds Obtained $t_2(ms)$ | Improvement $t_1/t_2$ |
|---|---|---|---|
| 10 | 15 | 0.89 | 16.85 |
| 20 | 63 | 6.90 | 9.13 |
| 30 | 140 | 28.50 | 4.91 |
| 40 | 256 | 82.30 | 3.11 |
| 50 | 546 | 187.10 | 2.92 |
| 60 | 1062 | 393.60 | 2.70 |

## 4.2 Experiments with SLM

Characterization of the SLMs is very important for optimal performance of the device. Both the SLMs used in the experiment were characterized and the compatibility of the SLMs for image projection and correction has been understood. SLM response at various parameters would help to locate an optimum range of operation.

The LC 2002 SLM has been characterized quite well earlier [16]. At brightness, b and contrast, c of 200, 128 respectively, maximum linearity of phase retardation has been shown. With this combination of brightness and contrast, a linear polarization angle, θ of 300 gives maximum width of grayscales where the retardance is linear. A set of plane images have been projected on the SLM and the total intensity was captured on the CCD. Here, plane images are the images with the same grayscale value assigned to all the pixels. The average grayscale value of the output images have been calculated for individual input plane images and were plotted. Fig. 9 shows the nonlinear behavior of the SLM, grayscale normalized to 1.

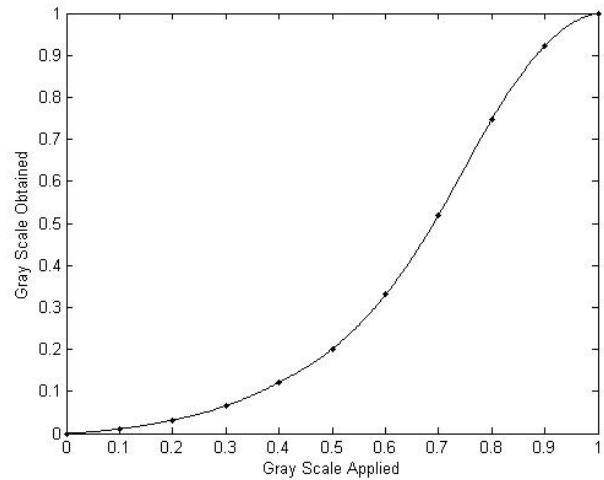

**Figure 9 Nonlinearity of the LC2002 SLM at b = 200, c = 128, *θ* = 300**

Meadowlark Hex127 SLM has nonlinearity as shown in the Fig. 10. It is possible to give either integer or half integer values of voltages to the different pixels of the Hex127 SLM, which will limit the number of gray scales that can be used. For example, if the voltage can be used from 3.5-7V, then we can expect a 3-bit gray scaling. So, a maximum range has been chosen for the SLM to give maximum phase change. This SLM is mostly nonlinear except in the range 3.5-9 V. This voltage corresponds to a phase change of nearly 1-2.5 radians.

The maximum phase shift that LC2002 SLM can offer is 1.25 radians. This phase retardance can be corrected using the Meadowlark SLM since it offers a maximum phase difference of 1.5 radians. The images were preprocessed for best performance and phase matching of the SLMs using inverse transformation.

**Table 2. Improvement in Strehl ratio after correction**

| Aberration Zernike polynomial | strehl ratio aberrated image | Corrected image |
|---|---|---|
| (1,1) | 0.5037 | 0.9795 |
| (2,0) | 0.1513 | 0.5977 |
| (2,2) | 0.3485 | 0.8979 |
| (3,1) | 0.2059 | 0.8358 |
| (3,3) | 0.3225 | 0.8719 |
| (4,0) | 0.4749 | 0.9914 |
| (4,2) | 0.3096 | 0.9038 |

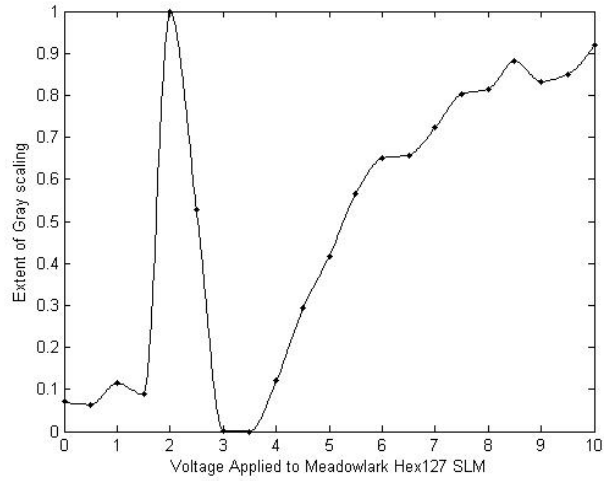

**Fig. 10. The nonlinearity of Meadowlark Hex127 SLM**

The aberrated and the corrected image are shown in Fig. 11 for a portion of the resolution chart image. The Strehl ratios have been computed and presented in Table 2.

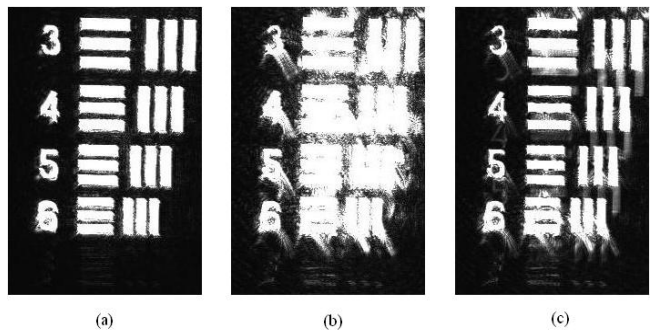

**Fig. 11. Aberrated and corrected images of a portion of the resolution chart image**

(a) Image without aberration (b) Aberration using $Z^0_2$ (c) Corrected image.

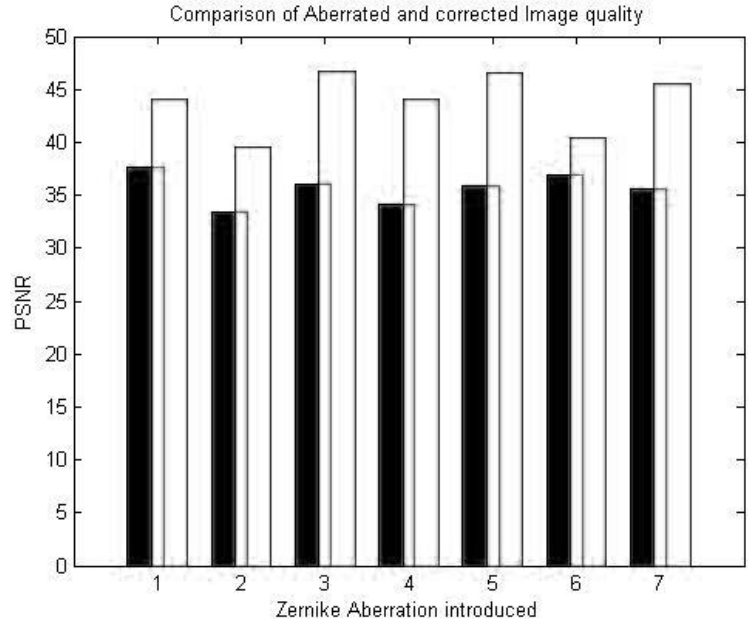

**Fig. 12. Comparison of PSNR before and after correction using resolution chart images**

Bar plot in Fig. 12 shows the improvement of the image quality after the correction.

## 5. CONCLUSIONS AND DISCUSSION

We detailed the generation and correction of Zernike aberrations imposed on an image using two SLMs. The possibility of using SLMs for phase corrections in an adaptive optics system has been highlighted. It is also possible to convert an amplitude and phase modulating SLM into a phase only SLM and thereby improve the phase modulation depth. The importance of the characterization of the SLMs for their nonlinearity and phase retardance has been stressed. Both the SLMs used in the experiment were characterized and are operated at optimum performance levels. The aberrations imposed on the images could be corrected to a great extent. The performance of the Meadowlark Optics Hex127 SLM as a wavefront corrector has been exemplified. The Strehl ratios were calculated and the improvement in the image quality has been quantified.

A fast method for the calculation of the Zernike moments suggested by Hosny has been confirmed [15]. Using this method and making further improvement in the algorithm, we have shown additional enhancement in quality and speed of the reconstructed image. The use of Zernike polynomials directly for phase restoration in an adaptive optics system has been suggested. We have also indicated the number of orders of polynomials required for reconstructing a phase image. The speeds at which the reconstruction can be done has been tabulated and plotted for various cases. Different applications can use these results for analysis of the aberrations. The accuracy of phase reconstruction depends on the number of Zernike polynomials used. Detailed analysis of the number of orders crucial for reconstruction process has been done and we conclude that for low resolution image reconstruction, less number of orders may be used for quick processing. For adaptive optics applications in astronomy, where the sensing is done at low resolution, we suggest using 15 orders of Zernike moments.

The highly reliable phase correcting ability of the SLM can make the device a very good wavefront corrector in closed loop adaptive optics systems.